\documentclass{emulateapj}
\accepted{Feb. 5 2007}
\slugcomment{}
\shorttitle{Variability of AGNs at z=0.36}
\shortauthors{Woo et al.}

\newcommand{\mbh}{M$_{\rm BH}$}

\newcommand{\kms}{${\rm kms}^{-1}$}

\begin{document}

\title{Variability of Moderate Luminosity Active Galactic Nuclei at z=0.36}

\author{Jong-Hak Woo\altaffilmark{1}, Tommaso Treu\altaffilmark{1},
Matthew A. Malkan\altaffilmark{2}, Matthew A. Ferry\altaffilmark{3},
Tony Misch\altaffilmark{4}}

\altaffiltext{1}{Department of Physics, University of California,
Santa Barbara, CA 93106-9530; woo@physics.ucsb.edu,
tt@physics.ucsb.edu} 

\altaffiltext{2}{Department of Physics and Astronomy, University of
California at Los Angeles, CA 90095-1547, malkan@astro.ucla.edu}

\altaffiltext{3}{Department of Physics, Mail code 103-33, California
Institute of Technology, CA 91125.}

\altaffiltext{4}{Lick Observatory, PO Box 85, Mt. Hamilton, CA 95140.} 

\begin{abstract}
We monitored 13 moderate luminosity active galactic nuclei at z=0.36
to measure flux variability, explore feasibility of reverberation
mapping, and determine uncertainties on estimating black hole mass
from single-epoch data. Spectra and images were obtained with
approximately weekly cadence for up to 4 months, using the KAST
spectrograph on the 3-m Shane Telescope. In broad band we detect
peak-to-peak variations of 9-37\% and rms variations of
2-10\%. The observed flux variability in the g' band (rest-frame
2800-4000\AA) is consistent with that in the r' band (rest-frame
4000-5200\AA), but with larger amplitude. 
However, after correcting for stellar light dilution,
using Hubble Space Telescope images, 
we find nuclear variability of 3-24\% (rms variation) 
with similar amplitudes in the g' and r' bands within the errors. 
Intrinsic flux variability of the H$\beta$ line is also
detected at the 3-13\% level, after accounting for systematic errors
on the spectrophotometry. This demonstrates that a reverberation
mapping campaign beyond the local universe can be carried out with a 3-m 
class telescope, provided that sufficiently long light curves are obtained. 
Finally, we compare the H$\beta$ FWHM measured from mean spectra with 
that measured from single-epoch data, 
and find no bias but an rms scatter of 
14\%,
mostly accounted for by the uncertainty on FWHM measurements.
The propagated uncertainty on black hole mass estimates, 
due to the FWHM measurement errors using
low S/N (10--15 per pixel) single-epoch spectra, is 30\%.
\end{abstract}


\section{Introduction}

The mass of supermassive black holes (\mbh) is the key parameter in
understanding the physics of Active Galactic Nuclei (AGNs) and the
role of black holes in galaxy formation and evolution, as indicated in
the local universe by the tight relations between \mbh\, and
host galaxy properties (Ferrarese \& Merritt 2000; Gebhardt et
al. 2000; Magorrian et al. 1998; Marconi \& Hunt 2003; Haring \& Rix
2004). However, direct dynamical measurements of \mbh\,
using spatially resolved kinematics of stars and gas around the
central black hole are limited to the local universe, where only about
three dozen masses have been measured (see Kormendy \&
Gebhardt 2001; Ferrarese \& Ford 2005; Tundo et al. 2006).

Beyond the local universe, \mbh\, can be estimated
for active galaxies using the so-called ``virial'' method. The method
assumes that the motion of the broad-line regions (BLR) is dominated
by the gravitational potential of the central source. Under this
assumption, \mbh\, can be determined from the dynamics of the BLR,
provided that velocity and size are measurable, and that some assumption on
the orbits is made in order to determine the virial coefficient. 
The velocity scale of the BLR can be estimated from the width of
broad-emission lines, while the size (R$_{\rm BLR}$) can be measured
using ``reverberation mapping'' (Blandford \& McKee 1982),
i.e. determining the time lag between continuum changes (presumably
originating from an accretion flow) and the corresponding changes in
the broad-emission lines. The ``virial'' assumption has been directly
tested for the best studied AGN, NGC 5548. Time lags for NGC 5548 have
been determined using several broad lines of different widths, and the
inferred distance from the central source shows the expected
``virial'' correlation for a central point mass, $V \propto R_{\rm
BLR}^{-1/2}$ (Peterson \& Wandel 2000).

Since the early attempts of determining the time lag between continuum
and broad line light curves with cross-correlation methods
(e.g. Gaskell \& Sparke 1986) and with more sophisticated response
function calculations (e.g. Krolik \& Done 1995), long-term efforts of
reverberation mapping of the local Seyfert galaxies (Wandel et
al. 1999) and the low-redshift ($z < 0.2$) PG quasars (Kaspi et
al. 2000) have provided clear detections of time lags for about three
dozens of highly variable AGNs (Peterson et al. 2004). The measured
time lags range from a few to $\sim$300 days and the nuclear optical
luminosities range between 10$^{43}$ and 10$^{46}$ erg s$^{-1}$. A
\mbh\, estimate is then obtained by combining the time-lags with
measures of the width of the broad lines and a virial coefficient. The
latter can be obtained for example by requiring that AGNs and their
host galaxies follow the same \mbh\,-velocity dispersion ($\sigma$; or
bulge luminosity) relation as quiescent galaxies (Onken et al. 2004;
Greene \& Ho 2006, Labita et al.\ 2006).  \mbh\, estimates via
reverberation mapping are believed to be accurate within a factor of
$\sim$3 (Bentz et al.\ 2006b; Onken et al. 2004; however, see Krolik
2001).

Reverberation mapping based masses have also provided a widely used
empirical method to estimate \mbh\, in distant AGNs from single-epoch
data, bypassing the need for observationally expensive time series.
This empirically calibrated photo-ionization method (Wandel et
al.\ 1999) is based on the observed correlation between the BLR size
(time lag) and optical luminosity at 5100\AA. The slope of the power
law relation R$_{\rm BLR}\propto L_{\rm 5100}^{\alpha}$ with
$\alpha\sim0.5$ is similar to that expected by photoionization models
of the BLR (Wandel et al. 1999; Kaspi et al. 2000; Bentz et
al. 2006a). This empirical relation has been widely used to determine
\mbh\, for large samples of broad-line AGNs, where long-term
monitoring efforts would have been prohibitive (e.g. Woo \& Urry 2002;
McLure et al. 2004). The uncertainty of \mbh\, estimates 
based on this empirical relation 
is believed to be 
a factor of 3-4 (Vestergaard 2002, 2006),
and it is due to several factors: i) intrinsic uncertainties in the
\mbh\, of local calibrators; ii) intrinsic scatter of the 
size-luminosity relation; iii) differences between single-epoch
measurements of optical luminosity and width, typically available for
high-redshift studies, and those obtained from mean or rms spectra
obtained from multiple epochs, generally used in the
reverberation mapping analysis.

Determining \mbh\, at high redshift is crucial to make
progress in a number of outstanding scientific issues, e.g., black
hole demographics (Yu \& Tremaine 2002), accretion mechanisms (Koratkar \& Blaes 1999), 
connections
between galaxy formation and evolution and AGN feedback (Croton 2006; Hopkins et al. 2006), origin
of the \mbh-$\sigma$ and other scaling relations (e.g. Silk \& Rees 1998; Kauffmann \& Haehnelt 2000). 
Studies of AGN variability at high-redshift and possibly direct reverberation
mapping determinations of time lags would be extremely valuable to
validate the photoionization method and understand its
uncertainty. However, reverberation mapping is challenging for bright
quasars at high-redshift because it requires a longer time base line
due to the time dilation $(1+z)$ effect. Moreover, high luminosity
objects (L$_{\rm 5100}$ $\geq$ 10$^{46}$ erg s$^{-1}$) seem to have
smaller amplitudes of continuum variability compared to Seyfert
galaxies (variability amplitude-luminosity anticorrelation; Cristiani
et al. 1997; Vanden Berk et al. 2004).
For example, after six year of monitoring, Kaspi et al. (2006)
detected no variability of the Ly$\alpha$ line for 6 high-redshift
(z=2.2-3.2) quasars while they detected C IV $\lambda$1550 variability
(see also Wilhite et al. 2006).
These high luminosity quasars show a factor of two smaller continuum variability 
than that of the lower luminosity PG quasars monitored
over a comparable rest-frame period.
Lower luminosity type-1 AGNs (Seyfert
1s), although fainter, provide several advantages from the point of
view of a monitoring campaign. First and foremost the expected time
lags are shorter (order of weeks to a few months), raising the hopes
of detecting a lag in only one season, without having to worry about
objects observability. Second, variability is expected to be more
pronounced since the amplitude of variability seems to be anticorrelated 
with luminosity.

In this paper we report on our time-domain study of 13 moderate
luminosity (L$_{\rm 5100}$ $\sim$ 10$^{44}$ erg s$^{-1}$) AGNs at z
$\sim$0.36. The campaign was carried out at the Lick 3-m telescope
from May 2004 to November 2004. The sample is drawn from our sample of
Seyferts, for which stellar velocity dispersion (Treu, Malkan \&
Blandford 2004; Woo et al.\ 2006; hereafter TMB04, W06) and bulge 
luminosity and radius
(Treu et al. 2006 in preparation) have been determined
from Keck spectroscopy and Hubble Space Telescope imaging in order to
investigate the cosmic evolution of the \mbh-$\sigma$ and \mbh-bulge
luminosity relations. The goals of this paper are to: i) study the
variability of distant seyferts to explore the feasibility of
reverberation mapping in the distant universe; ii) study the
variability of the H$\beta$ line width to determine the contribution
to the systematic uncertainty of the \mbh\, estimated
from single epoch spectra as opposed to rms or mean spectra. 
The paper is organized as follows, In \S~2 we describe observations and data
reduction. In \S~3 we describe continuum and line flux variability.
In \S~4 we compare line width measurements from single-epoch data.  In
\S~5 we discuss our results and their implications. A standard
cosmology is assumed where necessary (H$_0$=70 \kms
Mpc$^{-1}$, $\Omega_{\rm m}=0.3$ and $\Omega_{\Lambda}=0.7$).

\section{Observations and Data Reduction}

\subsection{Experiment Design and Sample Selection}

In order to increase our chances to detect a time lag in a single
season, we focused our efforts on moderate luminosity AGNs, for
which the time lag is predicted to be of order a few weeks in the rest
frame, based on the size-luminosity relation (Kaspi et al.\ 2005). A
continuos coverage is critical for a monitoring campaign, but the Lick
observatory does not offer service mode observations. Thus, we
decided to carry out our campaign over the summer months, 
when the typical weather patterns at Mt. Hamilton indicate 
that a relatively large fraction of time is useful for observations 
(above 80\% between June
and September\footnote{See http://mthamilton.ucolick.org/techdocs/MH\_weather/obstats/prcnt\_hrs.html.}).
Note that this program does not require photometric conditions, nor
particularly good seeing. Of course, the drawback of observing in the
summer is the short duration of the nights.
Given the expected time lags, we planned
observations with weekly cadence (corresponding approximately to 5
days in the rest frame). The Mt. Hamilton and Lick Observatory staff
helped enormously by scheduling with regular cadence, except for the
obviously longer gaps during full moon periods when the background was
prohibitive for optical observations.

The targets are drawn from a parent sample of 30 broad-line AGNs at
z=0.365$\pm$0.010, initially selected from the Sloan Digital Sky
Survey (SDSS) Data Release 2 (Abazajian et al. 2004) to investigate the relation
between \mbh\, and their host galaxy properties (for details
of sample selection, see W06). A subsample of 13 targets
was selected based primarily on observability during the summer
months, and secondarily on the availability of ancillary Keck and
Hubble data (Table~1).

 \subsection{Observations}      

All observations were performed with the Shane 3-m telescope at
the Lick Observatory between May 10 and November 19. Table~2 shows the
journal of observations.  The weather was generally bad in May and
after the end of September, effectively cutting our campaign to 4
months. A fraction of time was also lost to technical problems. When
we could open the dome, observing conditions ranged from photometric
to thick cirrus, with a typical seeing $\sim$2''. In total, more than
100 individual spectra and images were taken on 20 nights, allowing us
to measure variability of continuum and line fluxes.

We used the Kast double spectrograph with the 600 line mm$^{-1}$ grism
centered at $\sim$ 4300\AA~ to obtain blue (3300-5500\AA) spectra, and
with the 600 line mm$^{-1}$ grating centered at $\sim$ 7500\AA~ to
obtain red spectra (6300-9100\AA).  The pixel scale corresponds to
1.83 \AA$\times$0$\farcs$8 in the blue and 2.32\AA$\times$0$\farcs$8
in the red. The blue spectra cover the MgII line (rest wavelength
$\sim 2798$\AA) while the red spectra cover the H$\beta$ line (rest
wavelength 4861\AA) and extend in most cases to the red to
include the H$\alpha$ line. With the 2$''$ wide slit , the spectral resolution
(gaussian dispersion) measured from arc lines was $\sim$ 180 km
s$^{-1}$ around the MgII line and $\sim$ 100 km s$^{-1}$ around the
H$\beta$ line.  Typical exposure time for each object was 2700-3000
second or 2 $\times$ 1800 second, yielding S/N $\sim$ 10 per pixel on 
the continuum of single-epoch spectra.

Internal flat fields for the red spectra were obtained at each target
position to correct pixel-to-pixel variation and the fringing pattern
of the red CCD.  For the blue spectra, internal flats were taken in
the afternoon.  A set of A0V stars -- selected from the Hipparcos
catalog to be close in the sky to our targets -- was used to correct
the A and B-band atmospheric absorption features and to perform
secondary flux calibration (see TMB04 and W06 for
details). Spectrophotometric standards were observed for flux
calibrations.

For spectroscopic target acquisition and to obtain broad band imaging
time-series, we used the direct imaging capability of the Kast
spectrograph. Before every spectroscopic exposure, we obtained images
in the g' and r' bands over the unvignetted field of view of
approximately 2 arcminute diameter (0.8 arcsec pixel$^{-1}$).  The
field of view is large enough that for most objects at least a couple
of stars brighter than the target AGN were available for differential
photometry.  Typical exposure time for imaging was 120 second on the
red side with the r' filter and 150 second on the blue side with the
g' filter.

\subsection{Data Reduction}

We performed the standard data reduction including flat fielding,
wavelength calibration, spectral extraction, and flux calibration
using a series of IRAF scripts.  Cosmic rays were removed from each
exposure using the Laplacian cosmic-ray identification software (van
Dokkum 2001). Sky emission lines were used for wavelength calibration
in the red, supplemented by arc lamps in the blue.  One-dimensional
spectra were extracted for maximal S/N with a typical extraction
radius of 3 pixels, corresponding to $\sim$ 4.8$''$.
Thirteen low quality spectra with low S/N (mostly due to bad weather) or
severe fringing effects (due to imperfect flat-fielding) were removed
from the high quality sample to be analyzed for variability. 
The rest-frame mean and rms spectra for all 13 AGNs are shown in Figure~1,
together with the average noise level. Broad MgII, H$\beta$,
H$\alpha$ and narrow OIII lines are clearly visible in all the average
spectra. In most cases, the H$\beta$ line is not visible in the rms
spectra since the variation over the observed epochs is smaller than
the average uncertainty per pixel. 
The H$\alpha$ line is clearly visible in some of the rms spectra.

\section{Flux variability}

Variability is one of the main characteristics of AGNs, and is the
property that is relied upon for reverberation mapping
studies. However, the amplitude and the spectral shape change
is not well studied for general Seyfert galaxies and quasars.  
In this section, we first present optical continuum variability
(\S~\ref{ssec:contvar}) as determined by broad band g' and r'
photometry, using available HST photometry to separate the variable
nuclear component from the constant stellar component
(\S~\ref{ssec:intnucvar}). Then, in \S~\ref{ssec:specvar}, we study
the H$\beta$ line flux variability from the spectroscopic
analysis. We focus on the H$\beta$ line since this is the most commonly
used line in this redshift range, and the proximity of the narrow OIII
lines provides a robust relative spectrophotometric calibration.

\subsection{Continuum flux variability from Lick photometry}
\label{ssec:contvar}

In order to measure continuum variability, we performed differential
photometry using stars in the field of each AGN with both g' and r'
band images.  Three to five stars around the target AGN were typically
used to estimate relative flux variations of the AGN.  We excluded stars
at the edge of the images, where vignetting is problematic. Aperture
photometry of individual stars and AGNs was performed using the {\sc
phot} task in IRAF, with a 3 pixel (2.4$''$) aperture radius for flux
measurements and with an annulus between 8 and 12 pixels for sky
subtraction.  The optimal size of the aperture and sky subtraction
annulus were determined based on extensive tests with images of
various quality seeing.  Typical errors in the aperture photometry are
1-3\% for the 17-18th magnitude reference stars.

Magnitude differences between AGN and field stars were calculated and
then normalized to zero by subtracting the mean difference over all
observed epochs. Photometric errors of the target AGN and each star were added in
quadrature, yielding total errors on differential photometry in the
range 2-4\%. Normalized magnitude differences obtained with different
stars are consistent within the errors, providing a good sanity check
on the analysis.  The final light curve and errors were produced by
averaging normalized light curves and errors for individual
reference stars.

We measured the continuum flux variability for 7 objects, namely S04, S05,
S06, S24, S27, S40, and S99, with more than 5 epochs of reliable
observations (see Fig. 2, 3 and 4). A continuum broad band light
curve for S08 could not be obtained due to the lack of suitable
reference stars in the field. The observed flux variability in the g'
band, corresponding to rest-frame wavelength 2800-4000\AA, is
consistent with that of the r' band (rest-frame wavelength
4000-5200\AA), but with larger amplitude. In the g' band,
peak-to-peak variation is 12-37\% and rms variation is 2-10\% while in
the r' band peak-to-peak variation is 9-32\% and rms variation is
2-9\% (see Table~3). This amount of variability on week to month timescales
is consistent with previous studies of local Seyfert 1 galaxies and
quasars (e.g. Webb \& Malkan 2000; Klimek et al. 2004) and
high-redshift quasars (Kaspi et al. 2003). 

\subsection{Continuum nuclear flux and color variability}
\label{ssec:intnucvar}

Stellar light contributions from host galaxies to the total flux is
often significant, especially for Seyfert galaxies (Malkan \&
Filippenko 1983; Bentz et al. 2006a). This leads to reduced variability
in the integrated spectrum and hampers studying color
variability of the nucleus (Hawkins 2003). 

To investigate nuclear flux and color variability, we need to correct
for the host galaxy contamination. This can be achieved using the
AGN-galaxy decomposition analysis obtained from HST-ACS F775W (i')
band images (GO-10216; PI Treu; Treu et al.\ 2006, in preparation),
and adopting our single-epoch measurement of the nuclear light
fraction as our best estimate of the average nuclear light
fraction. First, we derived the stellar fraction for the g' and r'
bands using:
\begin{equation}
f_{m} = f_{i} \times 10^{-0.4[(m-i)_{gal}-(m-i)_{total}]}.
\end{equation}
Here, f$_{m}$ is the stellar fraction in each band, f$_{i}$ is the
stellar fraction in the i' band determined from the HST imaging
analysis, (m-i)$_{total}$ is the observed color from the SDSS
photometry (DR4), and (m-i)$_{gal}$ is the stellar light color,
$g-r=1.65\pm0.15$ and $r-i=0.60\pm0.06$, as estimated from population
synthesis models (Bruzual \& Charlot 2003).  
The uncertainty on the stellar colors represents the spread obtained from
models with ages ranging from 1 to 7 Gyr and metallicities
from 0.4 to 2.5 solar. Then, we derived intrinsic flux
variability by correcting the host galaxy contamination with:
\begin{equation}
R_{c} = {{R - f} \over {1 - f}}.
\end{equation}
Here, f is the stellar fraction in each band, R is the observed
variability ($R =F_{total}/<F_{total}>$), and R$_{c}$ is the corrected
variability ($R_{c}=F_{AGN}/<F_{AGN}>$). The mean correction on the variability
of 6 objects with HST images is 0.016 magnitude 
in the g' band and 0.057 magnitude in the r' band.
After correcting host galaxy contamination, we find 3-14\% rms variation 
in the g' band and 5-24\% in the r' band, somewhat larger in the red
for a few objects (see Table 3).

Figure 3 compares the continuum flux variability in the g' and r' bands
with and without host galaxy correction. 
Uncorrected flux variability
shows higher amplitude in the g' band than in the r' band ($\Delta
g'/\Delta r'=1.84^{+0.22}_{-0.18}$), as expected because stellar light
dilution is more prominent in the red. However the variations become
much more similar when the host galaxy contaminations are corrected
for ($\Delta g'/\Delta r'=0.71\pm0.06^{+0.13}_{-0.12}$,
where errors include a random component and a systematic error 
from uncertainties in stellar colors).
This suggests that the spectral shape in the rest-frame 2800-5200\AA~ 
does not significantly change on weekly time scales,
similar to the findings of Wilhite et al. (2005) 
that shows an ensemble quasar spectrum
is bluer in bright phases only at rest wavelength $<$ 2500\AA.
The slightly redder colors in bright phases are probably caused by
the lack of (or much lower) variability of emission lines, 
which are included in the wavelength coverage of the brodbands 
(MgII in blue, and H$\beta$ and [\ion{O}{3}] in red), because 
the relative emission line contribution to the continuum flux
is lower in bright phases, producing redder photometric colors
(Wilhite et al. 2005).

\subsection{Flux variability of the H$\beta$ line}
\label{ssec:specvar}

Emission line flux variability is more difficult to measure than
continuum variability, since the uncertainty in the flux calibration
of the ground-based spectroscopy is typically larger than ~10\%,
particularly for faint sources.  In the case of AGNs, however, narrow
emission lines such as [\ion{O}{3}] can be used as internal flux
calibrators, since these lines originate from extended low-density
regions -- probably more than 100 times further from the accretion
disk than the BLR (Bennert et al. 2006) -- and do not vary on short
time scales as broad-emission lines do.  Thus, flux variations of
broad lines can be measured by normalizing each spectrum to the
constant narrow line fluxes, eliminating uncertainties related to slit
loss effects, sky transparency, airmass correction, etc. An additional
advantage of the specific case of the doublet [\ion{O}{3}]$\lambda
\lambda$4959,5007 is that the flux ratio between the two components
does not vary (Bachall et al. 2004, Dimitrijevi{\'c} et al. 2007)
and therefore the rms of the measured ratio provides
a robust estimate of the residual uncertainties on the normalized flux.

For 8 objects (S04, S05, S06, S08, S24, S27, S40, S99) with more than
5 reliable single-epoch spectra, we measure H$\beta$ line flux
variations using [\ion{O}{3}] $\lambda$ 5007 line as an internal flux
calibrator as described below (the procedure is similar to that
adopted by local reverberation mapping studies, e.g. Peterson et
al. 2002). We also measure the [\ion{O}{3}] $\lambda$ 4959 to
[\ion{O}{3}] $\lambda$ 5007 line ratio to help optimize the wavelength
windows used for line flux determination\footnote{The optimal
wavelength windows for narrow line flux determination are a tradeoff
between the advantages of a small window -- high signal-to-noise per
pixel, little effects of continuum subtraction uncertainties -- and
the disadvantages -- sensitivity to uncertainties in wavelength
calibration as well as to pixelization in the wavelength domain. To
find the optimal window we repeated the procedure described below for
a range of wavelength regions and picked the one that minimizes the
scatter in the [\ion{O}{3}] doublet line flux ratio. The choice of the
wavelength region for the H$\beta$ line is far less critical since the decline
in S/N is more gradual and the spectrograph's pixels are small
compared to the line width.}.

First, we subtract the continuum under H$\beta$ and [\ion{O}{3}] lines
using the average flux around 4700\AA~ and 5100\AA~ with $\sim$ 50\AA~
windows.  We linearly interpolate these two continuum points to define
the underlying continuum.  Second, we measure the H$\beta$ and
[\ion{O}{3}] doublet line fluxes by integrating the flux over each
line in the optimal windows. Third, the H$\beta$ and [\ion{O}{3}]
$\lambda$ 4959 fluxes are normalized by the [\ion{O}{3}] $\lambda$
5007 flux.

Before we can proceed to determine the intrinsic H$\beta$ line flux
variability, we need to account for residual systematic errors in the
line ratios ($\sigma_{\rm sys}$). We use the fact that the flux ratio
of [\ion{O}{3}] $\lambda$ 5007 and 4959 is constant to obtain a
conservative estimate of $\sigma_{\rm sys}$. This is done by requiring
that a constant ratio be an acceptable fit to the data as measured by
the $\chi^{2}$ statistic:
\begin{equation}
\sum{ {{[R(i) - <R(i)>]^{2}} \over \sigma(i)^{2} + \sigma_{sys}^{2}} }= N-1.
\end{equation}
Here, R(i) is the flux ratio between [\ion{O}{3}] $\lambda$4959 and
$\lambda$5007 at each observed epoch, $\sigma(i)$ is the error in the
flux ratio, $\sigma_{sys}$ is the systematic error, and N is the
number of observed epochs. The inferred systematic errors are in the
range 2--15\%. We note that this is likely to be an upper
limit to the systematic errors as $\lambda$4959 is weaker than H$\beta$, 
and narrow lines fluxes are more sensitive to uncertainties on wavelength
calibration and to poor wavelength sampling than broad lines.

Taking random and systematic errors into account, we can now estimate
the H$\beta$ intrinsic line variability (Table 3). We find rms variability
in the range 3--13\%. We caution that this should be
considered a lower limit due to the fact that our estimate of systematic
errors is conservative. For 3 objects we did not detect significant
variability.

\subsection{Prospects for reverberation mapping}

We compare flux variations between continuum and the H$\beta$
line flux in Figure 4. The detected H$\beta$ line variability is
smaller than the rms variations of the continuum flux
(a factor of 2 smaller on average after host galaxy correction),
similar to other works (e.g. Rosenblatt \& Malkan 1992).  
Errors on the H$\beta$ flux measurements are larger than those on the
continuum flux measurements because the S/N of the spectra is
significantly lower than that of broad band images.  
The light curve of the H$\beta$ flux is qualitatively different from that of continuum
flux, possibly indicating a time lag. However, we do not detect a
clear lag between continuum and H$\beta$ light curves.  
This could be due to the fact that the time base line of our campaign is relatively short
(4 months at maximum) although the observed variation amplitude 
(5-10\% on H$\beta$)
is comparable to that of the time-lag measured samples in the literature.
To illustrate this, in Figure 5 we compared the monitoring time
base lines in the rest-frame of each object with the variation amplitude
for time-lag detected/undetected samples from the literature. In general
a few hundred days of time base lines were required to detect
a reliable lag. For high luminosity PG quasars, several years
of base lines were used to get a time-lag.
Out of 3 objects in our sample (open circles) 
with the longest time base line ($\sim$100 days in the rest-frame),
one object does not show H$\beta$ variability while the other two objects 
show comparable variability as the Seyfert galaxies with a time lag detection (filled squares).
In contrast, rest-frame time base line for our sample is at best in the borderline
of the regime, where a reliable time lag can be measured, suggesting that
we may detect a time lag if a factor of $\sim$2 longer time base line
can be provided. 

\section{H$\beta$ line width from single-epoch data}

The BLR velocity scale is one of the two ingredients in estimating
virial \mbh. In reverberation mapping \mbh\, estimates,
the FWHM of Balmer lines is generally measured from mean or rms
spectra averaged over many observed epochs (Kaspi et al. 2000).  In
contrast, for most of \mbh\, studies using the size-luminosity
relation, relatively low S/N data from a single-epoch observation,
e.g. SDSS quasar spectra, are used. This could introduce additional
uncertainty to virial \mbh\, estimates, if the width of broad
lines varies significantly or the random error on FWHM measurements on
single-epoch data is significant. 
Here, we investigate the uncertainty and variation
of the H$\beta$ line width by comparing FWHM from a mean spectrum with
that from each spectrum of various observed epochs. 

The FWHM of the H$\beta$ line is measured in several steps.
First, we subtract continuum under
the H$\beta$ line by identifying the continuum levels of each side of
the H$\beta$ line and interpolating between them.
Second, we subtract [\ion{O}{3}] $\lambda$4959 by dividing
[\ion{O}{3}] $\lambda$5007 by 3 and blueshifting. Third, we subtract
the narrow component of H$\beta$ by rescaling [\ion{O}{3}]
$\lambda$5007 and blueshifting it. 
Since the FWHM measurements are sensitive to the ratio of narrow H$\beta$ to [\ion{O}{3}]
(up to $\sim$10\%), we used the fixed ratio determined from the high S/N Keck spectra (see Table 3). 
Fourth, we fit the broad component
of H$\beta$ with the Gaussian-Hermite polynomials up to 5-6 orders,
depending on the asymmetry of the H$\beta$ line profile,
and measure the FWHM of the
model (Figure 6). This is necessary because our single-epoch spectra
have a typical S/N of 10-15, too low to precisely define
wavelengths at the half-maximum.

Figure 7 compares the H$\beta$ FWHM measured from mean spectra with that
measured from each single-epoch data for 9 objects with more than
3 reliable spectra.
S99 is excluded in this analysis since the H$\beta$ line has such a weak and
very broad profile that we could not get meaningful measurements from
single-epoch data. 
With 59 single-epoch data with S/N $\gtrsim10$, we find no systematic
bias with an average ratio of $1.03\pm0.02$ between single-epoch FWHM and
FWHM from mean spectra. 
The rms scatter around FWHM from mean spectra is $\sim$ 
14\%, similar to 15-20\% scatter
found by Vestergaard (2002), who compared one single-epoch FWHM with
FWHM from mean spectra for each of 18 AGNs.
This scatter is a combination of actual width variation over monthly timescales, and
random and systematic errors on single-epoch measurements.  
For example, Rosenblatt et al. (1992) measured H$\beta$ FWHM using 12 single-epoch spectra
for 12 local Seyfert galaxies and found 7-26\% rms scatter (mean rms scatter 
15\%),
which were comparable to their FWHM measurement uncertainty.

To understand intrinsic width variation and measurement errors,
we measured H$\beta$ FWHM for 8 objects (filled squares), 
for which high S/N (50-100) Keck single-epoch data are available (W06).
The random errors on the FWHM measurements from the Keck spectra 
are virtually negligible.
The rms scatter between the H$\beta$ FWHM from Lick mean spectra
and that from Keck single-epoch spectra is 
7\% (Figure 7). 
We note that this width variation is much smaller than continuum flux variation
as expected from the size-luminosity relation in which 
log FWHM scales with 1/4 $\times$ log L for given \mbh.
If we take this scatter as a measure of the intrinsic width variability,
then we can consider errors on the Lick single-epoch FWHM
to be 
12\% (0.05 dex), by subtracting 7\% (0.03 dex) variation from 14\%
(0.057 dex) scatter in quadrature.  Thus, most of the variation in the
single-epoch FWHM measurements seems to be caused by the measurement
uncertainty. Consequently, we do not find any correlation between
variations in FWHM and continuum or line flux since any intrinsic trend
is swamped by measurement errors.  If we take
14\% scatter as the uncertainty on FWHM measurements, the
propagated error in \mbh\, estimates based on single-epoch data
is 
30\%.  This estimate should be applied with caution to other
single-epoch data sets since the 30\% propagated error may not be
relevant because of the difference in S/N, the method of FWHM
determination, variability, and other factors.

\section{Discussion and Conclusions}

We detected variability of moderate luminosity ($L_{5100}\sim10^{44}$
erg s$^{-1}$) AGNs at z=0.36. Continuum flux variability in the
rest-frame wavelength 2800-4000\AA~ and 4000-5200\AA~ is detected at
the 2-10\% level (rms variation). These measurements are consistent
with other variability studies of local Seyfert galaxies and higher
redshift quasars (e.g. Webb \& Malkan 2000; Kaspi et al. 2000). 
After correcting for stellar light
contamination using HST-ACS images, we find that the intrinsic nuclear
continuum variability is 3-25\%. Blue and red continuum variability
correlates, indicating that the spectral shape does not significantly
change on weekly time scales.

We also detected intrinsic H$\beta$ line flux variability at the level of
3-13\%, on average a factor of 2 lower than the continuum flux variability. 
Light curves of the continuum flux and the H$\beta$ line flux are qualitatively different,
perhaps due to a time lag. However, the time base line of our campaign 
seems to be too short to detect a reliable lag.
We may detect a time lag if a longer time base line
can be provided. With modern scheduling techniques and perhaps a robotic telescope,
light curves could be extended to 8 months per year.

Finally we study variability of the H$\beta$ line width to estimate
uncertainties in \mbh\, estimates from single-epoch data. We
compare the H$\beta$ FWHM measured from high S/N mean spectrum with that
measured from each single-epoch spectrum. We find 
14\% scatter
around the one-to-one relationship, most of which comes from random
errors on FWHM measurements using single-epoch spectra.
\mbh\, estimated from single-epoch spectra with typical S/N of 10-15 
has 
30\% uncertainty due to the line width measurement errors. 
This uncertainty is significantly smaller than the
total uncertainty of \mbh\, estimates from single-epoch data (
a factor of 3-4; Vestergaard 2002), indicating
uncertainties on FWHM may not be a dominant source of uncertainties
on virial \mbh, estimates.


\acknowledgments

We thank the anonymous referee for a careful reading of the manuscript
and numerous suggestions that improved the presentation of our
results. We thank Aaron Barth, Roger Blandford, Omer Blaes and Brad
Peterson for numerous stimulating conversations. This work is based on
data collected at the Lick Observatory -- operated by the University
of California -- and with the Hubble Space Telescope operated by AURA
under contract from NASA. We are grateful to the Shane 3-m staff for
their help during the observations and Graeme Smith for help with the
scheduling. We thank Alex Filippenko and Ryan Foley for obtaining some
of the data for us. This project is made possible by the wonderful
public archive of the Sloan Digital Sky Survey.  We acknowledge
financial support by NASA through HST grant GO-10216.


\begin{deluxetable}{lcrrr}
\tablewidth{0pt}
\tablecaption{Targets}
\tablehead{
\colhead{(1) Name}        &
\colhead{(2) z}           &
\colhead{(3) RA (J2000)}     &
\colhead{(4) DEC (J2000)}    &
\colhead{(5) r'}    
}
\tablecolumns{5}
\startdata
S04 (J210211.50-064645.0) & 0.3580       &  21 02 11.51 &-06 46 45.03 & 18.75 \\
S05 (J210451.83-071209.4) & 0.3531       &  21 04 51.85 &-07 12 09.45 & 18.43 \\
S06 (J212034.18-064122.2) & 0.3689       &  21 20 34.19 &-06 41 22.24 & 18.53 \\
S08 (J235953.44-093655.6) & 0.3591       &  23 59 53.44 &-09 36 55.53 & 18.61 \\
S09 (J005916.10+153816.0) & 0.3548       &  00 59 16.11 &+15 38 16.08 & 18.33 \\
S12 (J021340.59+134756.0) & 0.3575       &  02 13 40.60 &+13 47 56.06 & 18.18 \\
S23 (J140016.65-010822.1) & 0.3515       &  14 00 16.66 &-01 08 22.19 & 18.22 \\
S24 (J140034.70+004733.3) & 0.3621       &  14 00 34.71 &+00 47 33.48 & 18.43 \\
S26 (J152922.24+592854.5) & 0.3691       &  15 29 22.26 &+59 28 54.56 & 18.93 \\
S27 (J153651.28+541442.6) & 0.3667       &  15 36 51.28 &+54 14 42.71 & 18.87 \\
S28 (J161156.29+451611.0) & 0.3682       &  16 11 56.30 &+45 16 11.04 & 18.78 \\
S40 (J012655.82+153357.8) & 0.3749       &   1 26 55.82 &+15 33 57.87 & 18.62 \\
S99 (MS 1558.3+4138)      & 0.3690       &  16 00 02.80 &+41 30 27.00 & 18.78 \\
\enddata
\label{T_target}
\tablecomments{
Col. (1): Target ID.  
Col. (2): Redshift.
Col. (3): RA.
Col. (4): DEC.
Col. (5): Extinction corrected $r'$ AB magnitude from SDSS photometry.
}
\end{deluxetable}

\begin{deluxetable}{lcrrrrrrrr}
\tablewidth{0pt}
\tablecaption{Journal of observations}
\tablehead{
\colhead{Run} &
\colhead{Date} &
\colhead{Targets} &
\colhead{Seeing}  &
\colhead{Conditions}     \\
\colhead{}       &
\colhead{}       &
\colhead{}           &
\colhead{arcsec}         &  
\colhead{}         \\
\colhead{(1)} &
\colhead{(2)} &
\colhead{(3)} &
\colhead{(4)} &
\colhead{(5)}}  
\tablecolumns{5}
\startdata
1 & 2004 May 10 &  S26,S99                            & 2.4-3    & clear/fog,humidity  \\
2 & 2004 May 18 &  S05,S23,S24,S27,S99                & 1.7-1.8  & clear  \\
3 & 2004 May 25 &  S23,S24,S26,S27,S28,S99            & 2.0-2.2  & photometric \\
4 & 2004 Jun 1  &  S05,S24,S26,S27,S28,S99            & 1.8-2.2  & photometric \\
5 & 2004 Jun 11 &  S05,S24,S26,S27,S28,S99            & 1.8-2.4  & cirrus \\
6 & 2004 Jun 20 &  S05,S24,S26,S27,S28,S99            & 2.1-2.8  & cirrus \\
7 & 2004 Jun 29 &  S05,S24,S26,S27,S28,S99            & 2.2-2.9  & clear  \\
8 & 2004 Jul 9  &  S24,S27,S99                        & 1.5-1.6  & clear/instrument failure  \\
9 & 2004 Jul 14 &  S05,S24,S27,S28,S99                & 2.2-2.4  & clear  \\
10 & 2004 Jul 18 &  S04,S05,S06,S27                    & 2-3      & cirrus \\
11 & 2004 Aug 6  &  S04,S05,S06,S08,S09,S27,S99        & 1.7-2.   & photometric/CCD problem \\
12 & 2004 Aug 14 &  S04,S05,S06,S08,S09,S27,S99        & 2.2-3    & clear/cirrus  \\
13 & 2004 Aug 24 &  S04,S05,S06,S08,S09,S27,S99        & 1.7-2    & cirrus \\
14 & 2004 Sep 8  &  S04,S05,S06,S08,S09,S12,S27,S40,S99& 1.3-1.7  & clear/guider problem  \\
15 & 2004 Sep 15 &  S04,S05,S06,S08,S09,S12,S27,S40,S99& 1.4-1.8  & photometric \\
16 & 2004 Sep 22 &  S04,S05,S08,S09,S12,S27,S40,S99    & 1.3-2.4  & cirrus \\
17 & 2004 Sep 30 &  S04,S05,S06,S08,S09,S27,S40        & 1.4-1.8  & cirrus \\
18 & 2004 Oct 11 &  S04,S05,S06,S08,S09,S12,S27,S40    & 1.8-2.2  & photometric/instrument problem \\
19 & 2004 Oct 20 &  S09,S40                            & 2-3      & cloud,humidity \\
20 & 2004 Nov 3  &                                     &          & rain   \\
21 & 2004 Nov 19 &  S12,S40                            & 2.5-2.7  & clear  \\
\enddata
\label{T_journal}
\tablecomments{
Col. (1): Observing run.
Col. (2): Observing date.
Col. (3): targets.   
Col. (4): Seeing (FWHM) measured from r' band direct imaging.
Col. (5): Conditions/Notes.
}
\end{deluxetable}

\thispagestyle{empty}
\begin{deluxetable}{lcrrrrrrrrrrrr}
\tablewidth{0pt}
\tablecaption{Variability}
\tablehead{
\colhead{Target}         &
\colhead{\# of epochs}  &
\colhead{$max_{g'}$}     &
\colhead{$rms_{g'}$}     &
\colhead{$e_{g'}$}      &
\colhead{$rms_{g' c}$}     &
\colhead{$max_{r'}$}     &
\colhead{$rms_{r'}$}     &
\colhead{$e_{r'}$}      &
\colhead{$rms_{r' c}$}     &
\colhead{rms([\ion{O}{3}])} &
\colhead{$\sigma_{sys.}$} &
\colhead{rms(H$\beta$)} &
\colhead{H$\beta$/[\ion{O}{3}]}  \\
\colhead{(1)} &
\colhead{(2)} &
\colhead{(3)} &
\colhead{(4)} &
\colhead{(5)} &
\colhead{(6)} &
\colhead{(7)} &
\colhead{(8)} &
\colhead{(9)} &
\colhead{(10)}&
\colhead{(11)}&
\colhead{(12)}&
\colhead{(13)}&
\colhead{(14)}}
\tablecolumns{14}
\startdata 
S04 &8/8/9    &0.338& 0.098 &0.026 & 0.118 &   0.096  &  0.022 &0.024 &  0.058 &   0.102 &0.068 &0.060 & 0.07\\
S05 &14/13/10 &0.121& 0.026 &0.026 & 0.029 &   0.111  &  0.028 &0.024 &  0.049 &   0.084 &0.051 &0.081 & 0.10\\
S06 &9/8 /5   &0.217& 0.061 &0.064 & 0.079 &   0.208  &  0.042 &0.043 &  0.127 &   0.014 &-     &-     & 0.10\\
S08 &0/0/8    &-    &-      &-     &-      &-         &-       &-     &-       &   0.200 &0.162 &-     & 0.10\\
S24 &6/7/6    &0.245& 0.078 &0.044 & 0.098 &   0.300  &  0.099 &0.037 &  0.243 &   0.111 &0.053 &0.049 & 0.08\\
S27 &16/17/15 &0.289& 0.078 &0.045 & 0.090 &   0.210  &  0.034 &0.033 &  0.075 &   0.165 &0.107 &-     & 0.18\\
S40$^{1}$ & 5/5/5   &0.147& 0.032 &0.057 &     - &   0.100  &  -     &0.054 &  -     &   0.035 &0.016  & 0.029 & 0.07 \\
S99 &14/15/11 &0.338& 0.072 &0.050 & 0.136 &   0.197  &  0.040 &0.028 &  0.193 &   0.066 &0.020 &0.130 & 0.10 \\
\enddata
\tablecomments{
Col. (1): Target ID.  
Col. (2): Number of flux points in the g' and r' band images and H$\beta$ line flux.
Col. (3): peak to peak variation in magnitude in g'
Col. (4): rms variation in g'
Col. (5): mean error in g'
Col. (6): rms variation in g' after stellar contribution correction. 
Col. (7): peak to peak variation in magnitude in r'
Col. (8): rms variation in r'
Col. (9): mean error in r'
Col. (10): rms variation in r' after stellar contribution correction.
Col. (11): rms scatter of the [\ion{O}{3}] $\lambda$5007 to [\ion{O}{3}] $\lambda$4959 flux ratio.
Col. (12): Systematic errors on H$\beta$ estimated as described in Section~3.3
Col. (13) Intrinsic rms H$\beta$ flux variation in magnitudes after correcting systematic errors.
Col. (14) narrow H$\beta$ component to [\ion{O}{3}] $\lambda$5007 flux ratio, used to remove narrow H$\beta$ component
before measuring broad H$\beta$ FWHM.
}
\tablerefs{
1) no HST image available. }
\end{deluxetable}

\clearpage

\begin{figure}
\epsscale{0.6}
\plotone{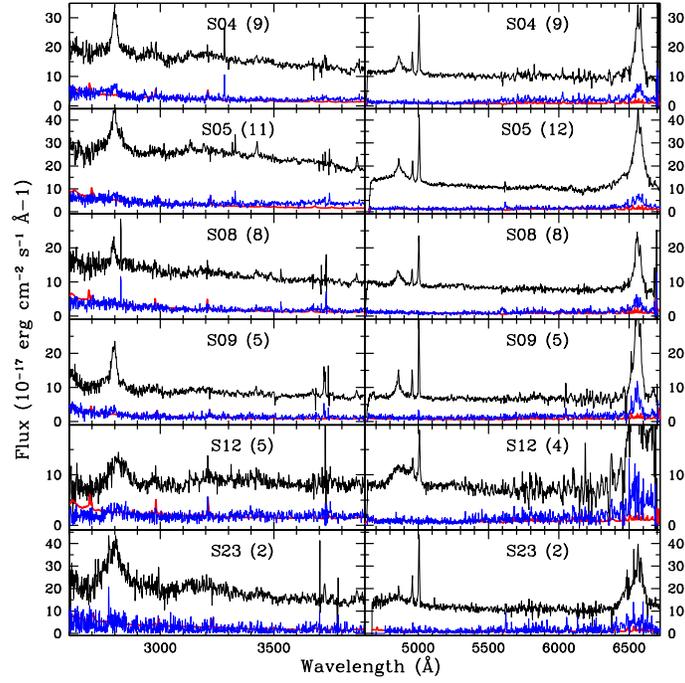}
\caption{Mean (black) and rms (blue) spectra of all observed AGNs, 
covering MgII, H$\beta$, and H$\alpha$ lines, with the 
average noise level (red).
The number of combined spectra is designated for each objects. 
For S06 and S99, blue spectra are excluded since MgII line
is not well defined due to the much lower S/N ratios.\label{fig:spectra}}
\end{figure}
\centerline{\includegraphics[width=.5\textwidth]{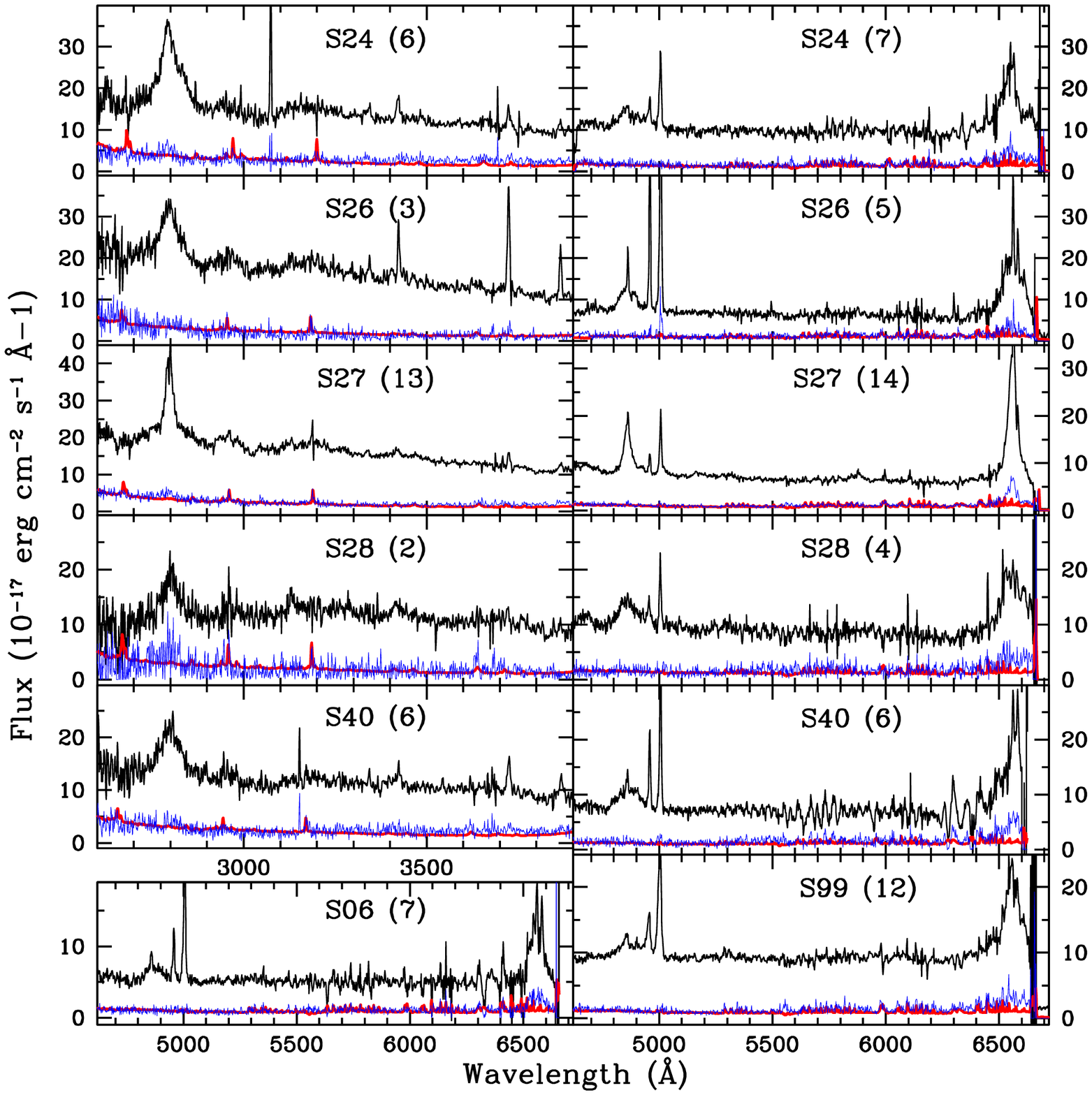}}
\vspace*{5mm}
\centerline{Fig. 1. --- Continued}
\clearpage

\begin{figure}
\epsscale{0.5}
\plotone{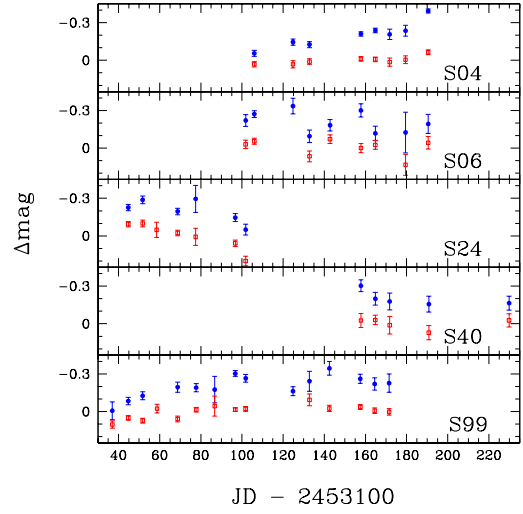}
\caption{Examples of light curves in the g' and r' bands.  Differential
photometry was obtained by comparison with nearby stars in the imaging
field as described in Section~3.1. For clarity, g' band light curves
(blue circles) are offset by 0.2 magnitudes.}
\end{figure}

\begin{figure}
\epsscale{0.8}
\plotone{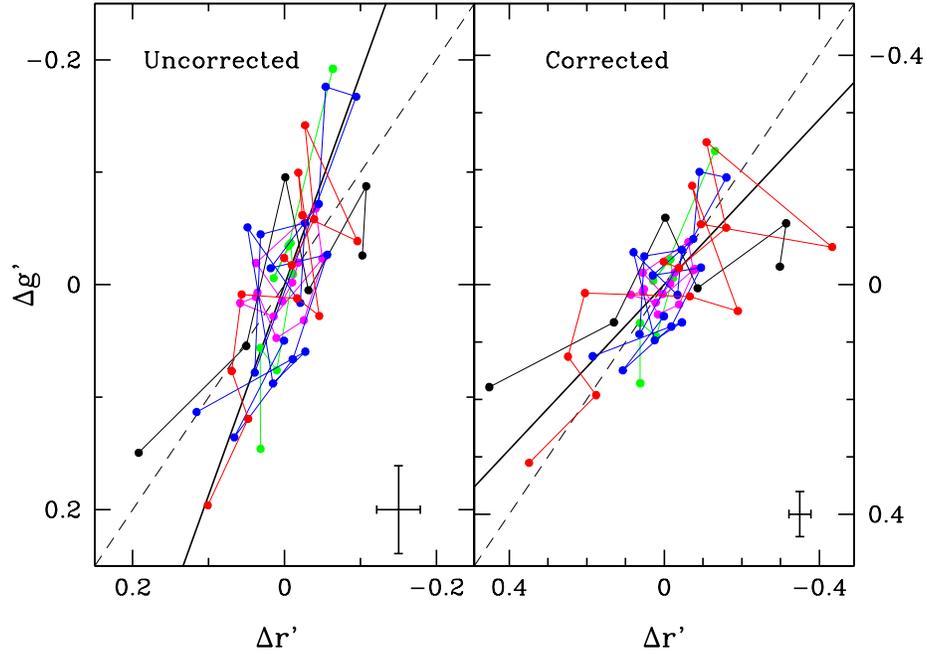}
\caption{Comparison of variability in g' and r' bands. Left: observed
flux variability. Individual AGNs are denoted with different colors
(green: S04, magenta: S05, yellow: S06, black: S24, blue: S27, red: S99).
Note that variations in g' and r' band are correlated,
with a larger amplitude in the blue band. The solid line is the best
fit slope, $1.84^{+0.22}_{-0.18}$, significantly larger than unity
(dashed line), suggesting that spectra look bluer when brighter.
Right: Nuclear flux variability after correcting for the stellar
contribution using HST-ACS images. Note that the amplitude of the
nuclear variations is larger and the plotting range is larger than in
the left panel. The best fit slope is now $0.71\pm0.06^{+0.13}_{-0.12}$,
where errors include a random component and a systematic error from uncertainties 
in stellar colors. This is marginally consistent with unity, 
suggesting approximately achromatic intrinsic luminosity variations 
over this wavelength range 2800-5200\AA.
Typical errors are plotted at the bottom-right of each panel.  
}
\end{figure}

\begin{figure}
\plottwo{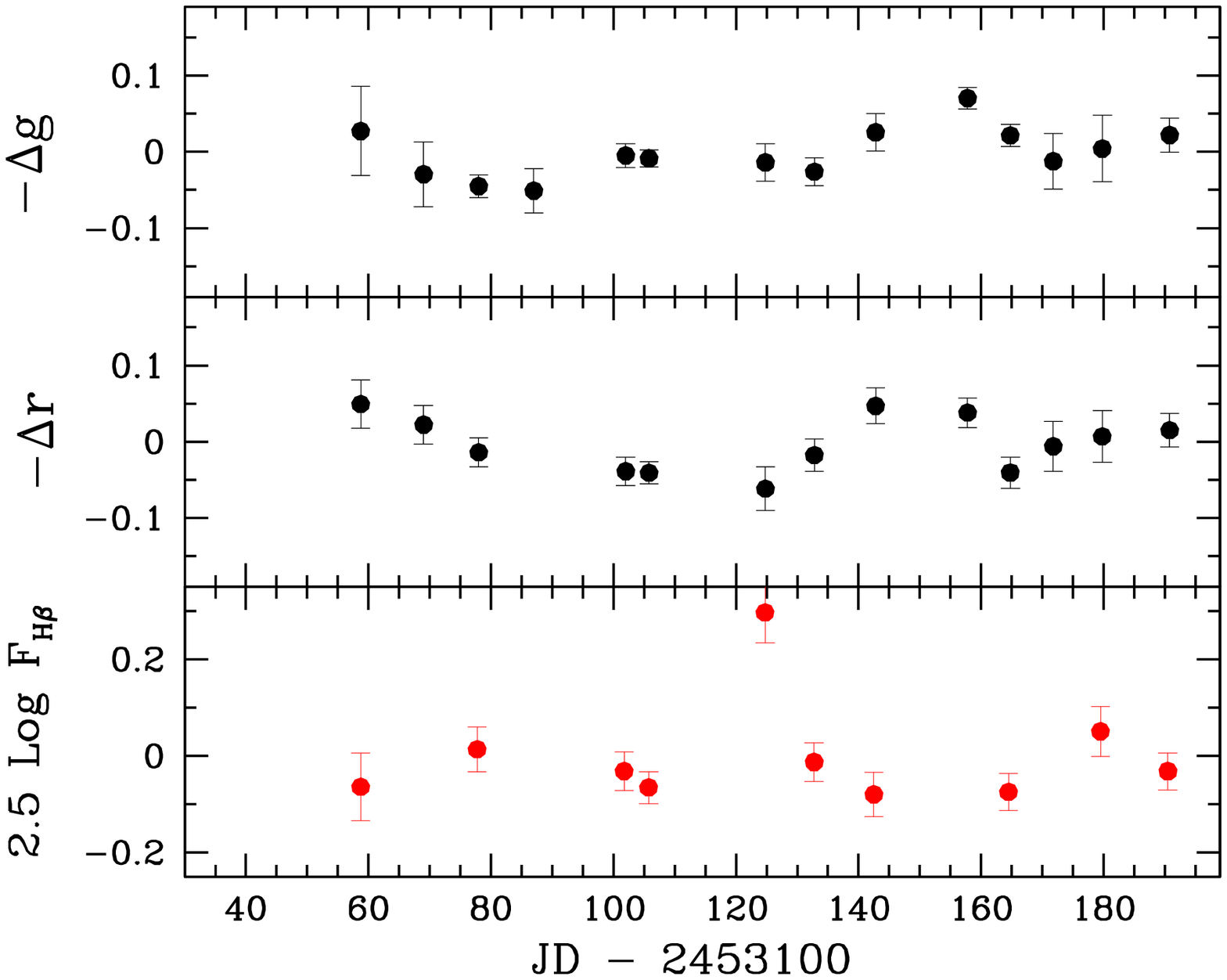}{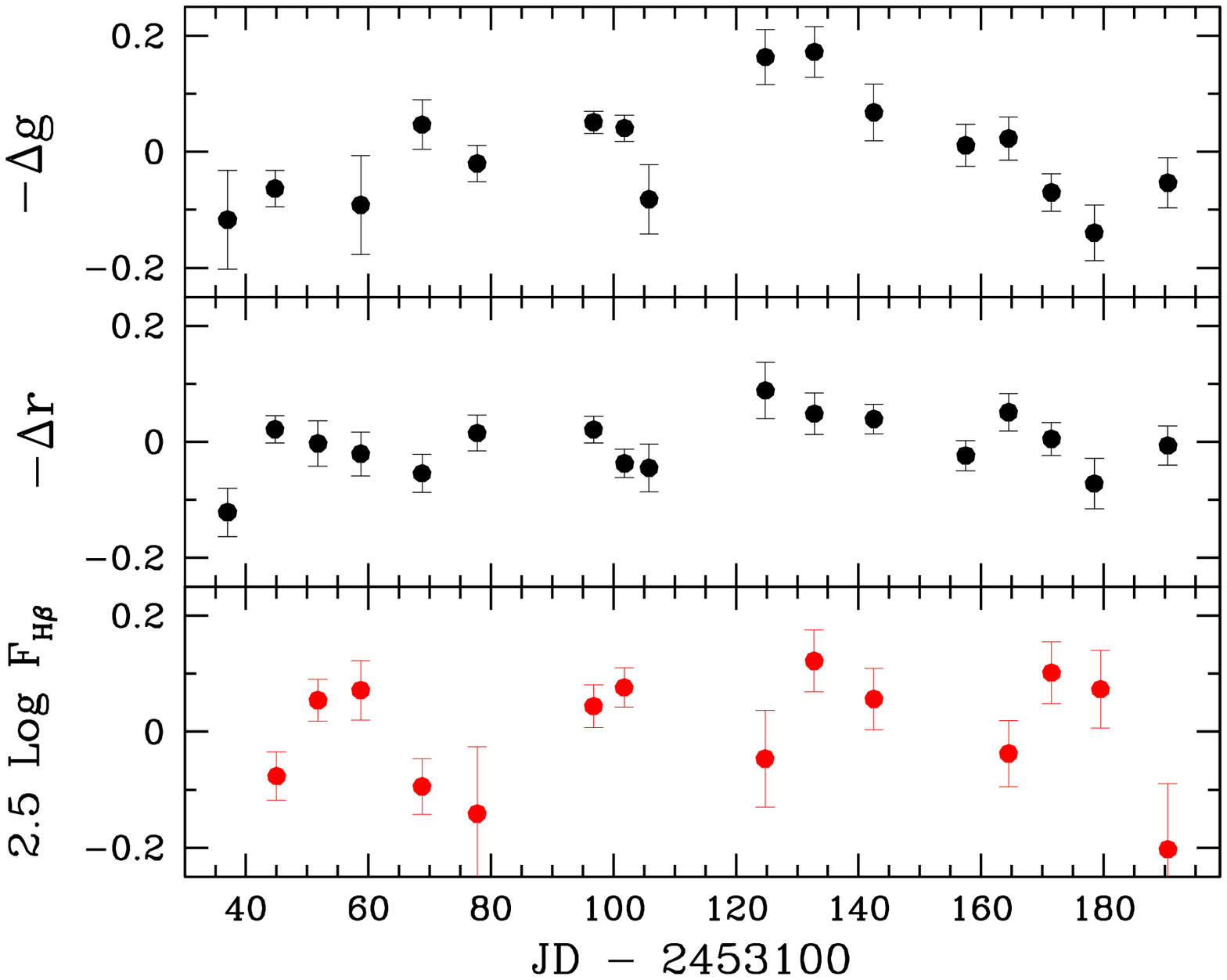}
\caption{Light curves of the continuum flux and the H$\beta$ line flux for
S05 and S27. For each object the continuum flux variability is shown in
the g' band (top) and the r' band (middle). 
The H$\beta$ line flux
light curve is shown in the bottom panel. Note that the amplitude of
the variations is similar to that of the broad band.  }
\end{figure}

\begin{figure}
\epsscale{0.7}
\plotone{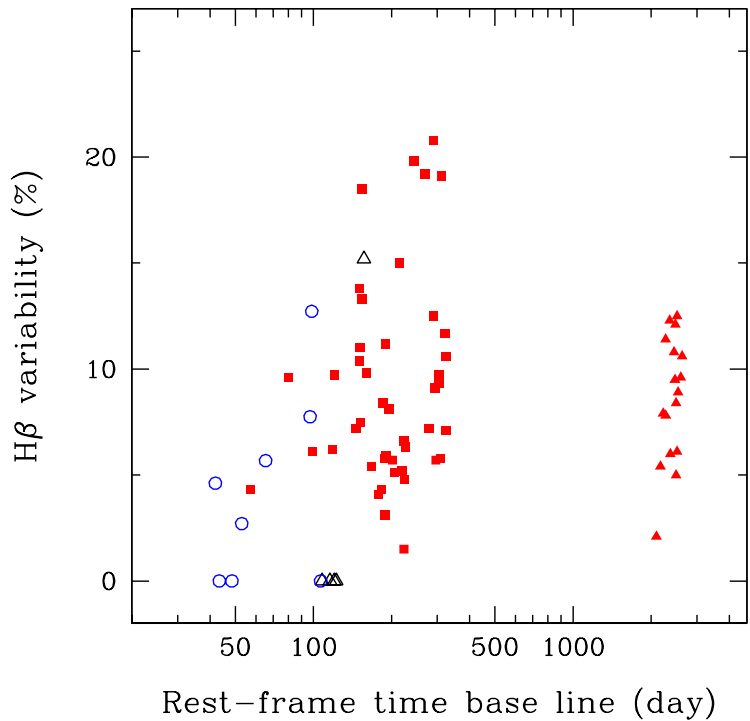}
\caption{H$\beta$ variability vs. rest-frame time base line for various samples.
Time lag detected Seyferts (filled squares) and PG quasars (filled triangles)
are from the compilation of Peterson et al. (2004).
AGNs with no or unreliable time lag detection are denoted with open symbols
(circles: our Seyferts at z=0.4; triangles: 5 AGNs from Lovers of AGNs 
campaign, Jackson et al. 1992; Dietrich et al. 1994; Stirpe et al. 1994; Erkens et al. 1995).
Time base line for time lag measured sample is generally more than a few hundreds days 
in the rest-frame. In contrast, time base line for our sample is relatively short
($<$100 days) although variability of the H$\beta$ line flux is comparable to
that of the time-lag measured Seyferts.
}
\label{time}      
\end{figure}

\begin{figure}
\epsscale{0.5}
\plotone{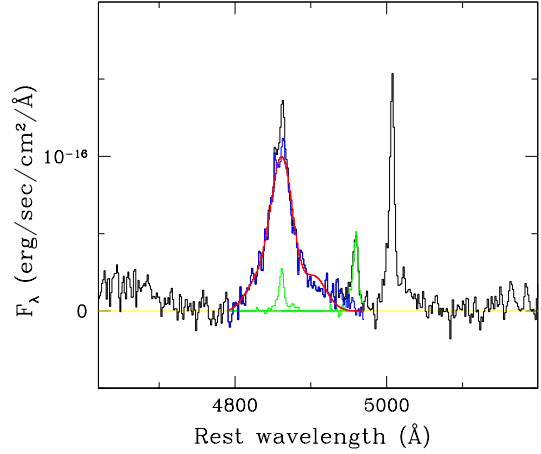}
\caption{Example of the H$\beta$ FWHM measurement using Gaussian-Hermite
polynomials. After removing the narrow H$\beta$ and [\ion{O}{3}]
$\lambda$4959 with a rescaled and blueshifted [\ion{O}{3}
$\lambda$5007 line profile (green), Gaussian-Hermitian models
(red) are used to fit the broad H$\beta$ line (blue). 
The FWHM of the line is then measured on the model fit.}
\label{FWHM_fit}      
\end{figure}

\begin{figure}
\epsscale{0.7}
\plotone{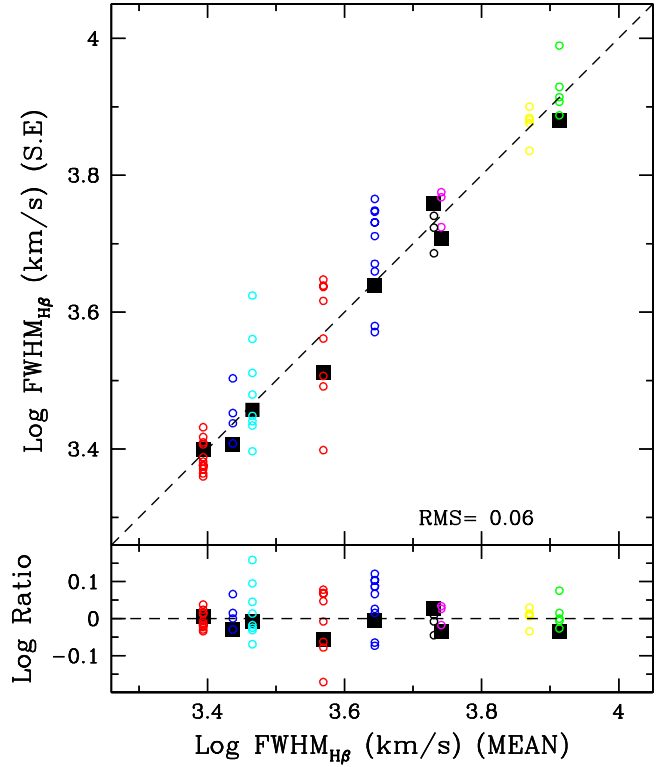}
\caption{Comparison of the H$\beta$ FWHM measured from mean spectra with
those measured from various single-epoch spectra. Single-epoch FWHMs
are consistent with FWHM from mean spectra with 
an average ratio of $1.03\pm0.02$.
The rms scatter around FWHM from mean spectra is
14\%,
indicating uncertainties of \mbh\, estimates based on single-epoch data can be 
30\%
due to the random errors in measuring FWHM. For comparison, we show   
FWHM measurements using high S/N Keck data (filled squares),
which are consistent with FWHM of mean spectra with a rms scatter
$\sim$0.03 dex, indicating intrinsic FWHM variation is $\sim$7\%.}
\label{com_FWHM}
\end{figure}

\end{document}